\documentstyle[12pt]{article} 
\def\F{{\bf F}}

\def\beq{\begin{equation}}
\def\eeq{\end{equation}}
\def\pl{\partial}

\def\al{\alpha}
\def\bt{\beta}

\def\ga{\gamma}
\def\de{\delta}
\def\De{\Delta}

\def\Si{\Sigma}
\def\te{\theta}

\def\lam{\lambda}

\def\ep{\epsilon}

\def\l{\left (}
\def\r{\right )}
\def\fr{\frac}
\def\la{\label}
\def\hs{\hspace}
\def\vs{\vspace}

\def\ran{\rangle}
\def\lan{\langle}
\def\ov{\overline}
\def\tl{\tilde}
\def\tm{\times}

\begin{document}

\begin{titlepage}
\begin{flushright}
BA-03-04\\
\end{flushright}

\begin{center}
{\Large\bf    
Neutrino Democracy And Other Phenomenology 

From 5D $SO(10)$}
\end{center}
\vspace{0.5cm}
\begin{center}
{\large 
{}~Qaisar Shafi$^{a}$
\footnote {E-mail address: shafi@bartol.udel.edu},~ 
{}~Zurab Tavartkiladze$^{b, c}$
\footnote {E-mail address: Z.Tavartkiladze@ThPhys.Uni-Heidelberg.DE} 
}
\vspace{0.5cm}

$^a${\em Bartol Research Institute, University of Delaware,
Newark, DE 19716, USA \\

$^b$ Institute for Theoretical Physics, Heidelberg University, 
Philosophenweg 16, \\
D-69120 Heidelberg, Germany\\

$^c$ Institute of Physics,
Georgian Academy of Sciences, Tbilisi 380077, Georgia\\
}

\end{center}
\vspace{1.0cm}

\begin{abstract}

We present a five dimensional supersymmetric $SO(10)$ model 
compactified on an orbifold $S^{(1)}/Z_2\tm Z_2'$. 
The gauge symmetry
$G_{422}\equiv SU(4)_c\tm SU(2)_L\tm SU(2)_R$,  
realized on one of the fixed points (branes), is spontaneously broken to 
the MSSM via the higgs mechanism. Employing a flavor ${\cal U}(1)$ 
symmetry and suitably extending the 'matter' sector enables 
us to understand large mixings in the neutrino sector via a democratic 
approach, versus the small CKM mixings. 
A residual ${\cal R}$-symmetry on $G_{422}$ brane helps eliminate the 
troublesome dimension five nucleon decay,
while the ${\cal U}(1)$ symmetry plays an essential role
in suppressing dimension six decay. 
For rare leptonic decays we expect 
${\rm BR}(\mu \to e\ga)\sim {\rm BR}(\tau \to \mu \ga)\sim
{\rm BR}(\tau \to e\ga)$.

\end{abstract}

\end{titlepage}

\section{Introduction}

The recent SuperKamiokande \cite{atm}, \cite{sol} and KamLAND 
\cite{kamland}
data have provided increasingly strong evidence for
oscillations between active neutrino flavors. Namely,
the atmospheric neutrino deficit is due to (near) maximal 
$\nu_{\mu }\to \nu_{\tau }$ oscillations with 
\beq
\sin^2 2\te_{\mu \tau }\simeq 1~,~~~~~ 
\De m^2_{\rm atm}\simeq 3\cdot 10^{-3}~{\rm eV}^2~,  
\la{atmdata}
\eeq
while the preferred solution for the solar neutrino
anomaly is the large mixing angle (LMA) solution with 
$\nu_e\to \nu_{\mu , \tau }$ oscillations, such that
\beq 
\sin^2 2\te_{e \mu, \tau }\approx 0.8~,~~~~ 
\De m^2_{\rm sol}\sim 10^{-4}~{\rm eV}^2~.
\la{soldata}
\eeq 
These developments require a physics beyond the Standard Model (SM). The
most popular extension is the introduction of right handed (SM
singlet) neutrinos with superheavy Majorana masses, which 
induce suppressed neutrino masses through the see-saw mechanism
\cite{seesaw}. A consistent framework for the solution of the gauge 
hierarchy problem is provided by supersymmetry which leads one to think 
that a supersymmetric GUT theories such as $SO(10)$ 
\cite{firstSO10} are quite compelling. 
In building
realistic models, in addition to the neutrino data, one should
also take into account the charged fermion sector and see how these
two sectors blend together. The three quark-lepton
families display a clear intergeneration hierarchy. Only the top
quark mass [$m_t=(174.3\pm 5.1)$~GeV] is close to the electro-weak 
symmetry
breaking scale, which means that the top Yukawa coupling 
$\lam_t\simeq 1$. 
Introducing the parameter
$\ep \simeq 0.2$ (close in value to the Cabibbo angle $V_{us}$) one can
express the hierarchies between the Yukawa couplings as follows:
\beq
\lambda_t\sim 1~,~~
\lambda_u :\lambda_c :\lambda_t \sim
\epsilon^8:\epsilon^4 :1~,
\la{ulam}
\eeq
\beq
\lam_b\sim \lam_{\tau}~,~~~
\lambda_d :\lambda_s :\lambda_b \sim
\epsilon^4:\epsilon^2 :1~,
\label{dlam}
\eeq
\beq
\lambda_e :\lambda_{\mu } :\lambda_{\tau } \sim
\epsilon^5:\epsilon^2 :1~.
\label{elam}
\eeq
As far as the CKM mixing angles are concerned, they have relatively 
small values
\beq
V_{us}\sim \epsilon~,~~~V_{cb}\sim \epsilon^2~,~~~
V_{ub}\sim \epsilon^3~.
\label{ckm}
\eeq

It is a challenging task to provide a compelling
framework for understanding the origin of the small CKM mixing
angles in (\ref{ckm}) versus the large  neutrino mixings. The
hierarchies in (\ref{ulam})-(\ref{elam}) also should be 
adequately understood. Some flavor symmetry is well motivated in this 
context, and one simple choice is provided by an abelian ${\cal U}(1)$
symmetry, which can be successful in building the desired
charged fermion \cite{feru1}, \cite{1feru1} and neutrino 
\cite{nuu1MSSM}-\cite{nuu1flipSU6} sectors. With a
${\cal U}(1)$ symmetry, the variety of textures for neutrino mass 
matrices can be constructed, giving either maximal or large neutrino 
mixings. This
has been pursued within $\nu $MSSM  \cite{nuu1MSSM}, meaning that MSSM is 
augmented by right handed neutrinos 
(models such as $SO(10)$ \cite{firstSO10}, 
$G_{422}\equiv SU(4)_c\tm SU(2)_L\tm SU(2)_R$ \cite{PSmodel} etc. 
automatically introduce right handed neutrinos)
and ${\cal U}(1)$ flavor symmetry, and
in some cases yields interesting predictions. The same 
approach has been used in realistic
GUTs such as $SU(5)$ \cite{nuu1SU5}, $SO(10)$ \cite{nuu1SO10},   
$G_{422}$ \cite{our422}, 
flipped $SU(5)$ \cite{ourdem},
pseudo-Goldstone $SU(6)$ \cite{nuu1SU6} and $E_6$ \cite{nuu1E6}.
A ${\cal U}(1)$ symmetry was also used within $SU(6)\tm U(1)$ 
\cite{nuu1flipSU6}, $SU(3)_c\tm SU(3)_L\tm SU(3)_R$ \cite{nuu1SU3cub} and
$SU(6)\tm SU(2)$ \cite{nuu1SU6SU2}\footnote{These three
groups  can emerge through $E_6$ breaking \cite{SU3fromE6}, 
\cite{flSU6fromE6}  
and have interesting phenomenological implications.}.

An interesting possibility, different from models
relying on specific textures to
obtain large neutrino mixings is the so-called {\it democratic} approach 
\cite{dem}, \cite{ourdem}, \cite{5Dourdem}, in which all
flavors of left
handed lepton doublets $l_{\al }$ ($\al =1, 2, 3 $ is a generation index) 
transform identically under ${\cal U}(1)$. That is, the flavor symmetry does
not distinguish neutrino flavors and consequently one naturally expects
large mixings between them. At the same time, the ${\cal U}(1)$ 
symmetry is still
crucial for understanding hierarchies between charged fermion masses and
the CKM mixing angles. It was shown in \cite{ourdem} that within 
the MSSM, neutrino
democracy nicely blends with the charged fermion sector, but within $SU(5)$
one encounters difficulties. Namely, with ${\cal U}(1)$ flavor symmetry and
neutrino democracy, $SU(5)$ GUT with minimal matter content gives
unacceptably small $V_{us}$($\sim \ep^3\sim 1/125$).
The root of this problem is in unified $SU(5)$ multiplets where the
quark-lepton states come from. Similar difficulty can also be expected  
in other GUTs with matter states embedded in a single GUT 
representation. 
It is possible to avoid this problem either by considering
extended GUTs such as flipped $SU(5)$ \cite{ourdem} or constructing
$SU(5)$
GUT in five dimensions \cite{5Dourdem}. With a minimal matter 
sector these theories typically yield unsatisfactory 
Yukawa matrices, such as
$\hat{M}_d^{(0)}=\hat{M}_e^{(0)}$ in $SU(5)$, and
$\hat{M}_u^{(0)}\propto \hat{M}_d^{(0)}=\hat{M}_e^{(0)}$, 
$\hat{V}_{CKM}={\bf 1}$ in $SO(10)$. 
This indicates the need for an extended matter sector as one possible way
to resolve these problems.

We also note that orbifold constructions \cite{kawa}-\cite{5Dso10}
turn out to be very efficient and powerful in resolving GUT problems
such as doublet-triplet (DT) splitting, nucleon stability, 
unwanted asymptotic relations,
GUT symmetry
breaking etc. For earlier attempt within superstring
derived models see for instance ref. \cite{stringDT}. 
In addition, the five dimensional
setting offers an economical way to realize neutrino democracy as well as
the desired charged fermion pattern \cite{5Dourdem}. 
Within the orbifold construction, different 
ways for realizing large neutrino mixings 
in the presence of ${\cal U}(1)$ symmetry were considered in
\cite{highD}, \cite{5Dso10}. The  scenarios with  democratic approach have 
some peculiar features which can be tested in future experiments. Namely,
models with democracy between $l_{\al }$ states  
predict that BR$(\mu \to e\ga )\sim $BR$(\tau \to \mu \gamma)\sim $BR
$(\tau \to e\gamma)$. 
Also, SUSY GUT democratic
scenarios yield nucleon decay with emission of charged leptons 
$e, \mu$ with comparable partial lifetimes 
$\tau (p\to Ke)\sim \tau (p\to K\mu )$. This would distinguish them
from models in which the process $p\to K\mu $ dominates.

In this paper we present a 5D SUSY $SO(10)$ GUT augmented with ${\cal U}(1)$
flavor symmetry. By orbifold compactification, on one of the fixed points
we realize the gauge symmetry $G_{422}$ 
\cite{bran422}\footnote{Obtaining the $G_{422}$ model from higher 
dimensional $SU(4)_c\tm SU(4)_{L+R}$ and $SO(12)$ GUTs was discussed in
\cite{mu2}, \cite{ilo}}.  
To implement neutrino democracy and obtain realistic charged 
fermion masses, some
additional matter states play a crucial role. They can be introduced
either in the bulk or directly on the fixed point
(so, this model differs from a recently suggested scenario 
\cite{5Dso10}, 
where an abelian flavor symmetry also plays an important role). 
Once this extension is done, we find that neutrino democracy 
is an automatic consequence of the model. Due to 
left-right
symmetry at the fixed point, the mass of the lightest right handed 
neutrino is predicted to be
$\sim 6\cdot 10^{10}\tan^2\bt$~GeV ($\tan \bt $ is a ratio of
VEVs of the MSSM higgs doublets).  This scale can be interesting for
understanding the observed baryon asymmetry either through thermal 
or inflationary leptogenesis \cite{Lgen}, \cite{infry}, \cite{sacha}.
The question of nucleon decay is also addressed and it is shown 
that under certain conditions the nucleon can be
absolutely stabile. 
This may be answer to the question of why the search for proton decay 
has been unsuccessful (at least so far!). 
In some other schemes that we consider, dimension six proton decay may be
accessible in future experiments. 
We also discuss rare leptonic decays and some other phenomenological
consequences.

\section{5D SUSY $SO(10)$ On An $S^{(1)}/Z_2\tm Z_2'$ Orbifold}

We consider a supersymmetric (SUSY) 
$SO(10)$ gauge theory in five dimensions (5D) compactified on an
$S^{(1)}/Z_2\tm Z_2'$ orbifold.
In terms of 4D superfields, 5D $N=1$ SUSY
multiplets correspond to 4D $N=2$ supermultiplets. The gauge superfield
is $V_{N=2}=(V, \Si )$, where $V$ and $\Si $ are vector and chiral
superfields respectively, transforming in the adjoint of $SO(10)$. 
For bulk matter and higgs superfields, each chiral state
$\Phi $ is accompanied by its mirror $\ov{\Phi }$, so that they constitute 
an $N=2$ supermultiplet ${\bf \Phi}_{N=2}=(\Phi , \ov{\Phi })$.
By compactification on $S^{(1)}/Z_2\tm Z_2'$, it is possible
to break $SO(10)\to G_{422}$ on one of the fixed points. 
In terms of $G_{422}$, an adjoint ${\bf 45}$ of $SO(10)$ reads
\beq
{\bf 45}=L(1, 3, 1)+R(1, 1, 3)+C(15, 1, 1)+B(6, 2, 2)~,
\la{45dec}
\eeq
where the transformation properties under $G_{422}$ are indicated. 

The fifth
space-like coordinate $y$ of 5D parameterizes a compact circle 
$S^{(1)}$ with
radius $R$, and the orbifold parities act as follows:
$Z_2:~y\to -y$, $Z_2':~ y'\to -y'$ ($y'=y+\pi R/2$). Under $Z_2\tm Z_2'$,
the bulk states have definite parities 
$(P, P')=(+, +), (+, -), (-, +), (-, -)$. The KK masses of the corresponding
states are 
\beq
2n \mu_0~,~~ (2n+1)\mu_0~,~~ (2n+1)\mu_0~,~~ (2n+2)\mu_0~,
\la{KKmass}
\eeq
where $\mu_0=1/R$ and $n=0, 1, 2, \cdots $ 
denotes the quantum number of the KK state in
mode expansion. 
{}From (\ref{KKmass}) one observes that only states with 
$(+, +)$ parity contain zero modes, while the remaining states  
acquire masses $\sim \mu_0 $.

The various components from $V(45)$, $\Si (45)$ are assigned the following
$Z_2\tm Z_2'$ parities:
$$
(V_C, V_L, V_R)\sim (+, +)~,~~~~~V_B\sim (-, +)~,
$$
\beq
(\Si_C, \Si_L, \Si_R)\sim (-, -)~,~~~~~\Si_B\sim (+, -)~.
\la{VSipar}
\eeq
At the fixed point $y=0$, we have the residual gauge symmetry
$G_{422}$, while at the fixed point $\pi R/2$ we find a $SO(10)$ 
symmetry.
Both at $y=0$ and $y=\pi R/2$, we also have 4D $N=1$ SUSY 
leftover from the original 5D $N=1$ SUSY.

In
order to build the fermion sector at $y=0$ and arrange for $G_{422}$
breaking down to the MSSM, we should introduce matter and
higgs superfields. It turns out that by introduction of appropriate states
in the bulk, it is possible to obtain the desired zero mode representations
of $G_{422}$ \cite{PSmodel}. 
For this, one needs to 
introduce additional states \cite{arch}, the so-called 
'copies'  (see $1^{\rm st}$ and $3^{\rm rd}$ in ref. \cite{symbr})
in the bulk. Namely, we introduce
${\bf 16}^{F}_{N=2}=(16^F, \ov{16}^F)$ and
${\bf 16}^{F^c}_{N=2}=(16^{F^c}, \ov{16}^{F^c})$ (for each
generation) where, under $G_{422}$
\beq
16^F=F(4, 2, 1)+F^{c\hs{0.5mm}'}(\bar 4, 1, 2)~,~~
16^{F^c}=F^{{}\hs{0.5mm}'}(4, 2, 1)+F^c(\bar 4, 1, 2)~,
\la{16dec}
\eeq
and similarly for $\ov{16}^F$, $\ov{16}^{F^c}$.
With the parity prescriptions
\beq
(F, F^c)\sim (+, +)~,~~~~(F^{{}\hs{0.5mm}'}, 
F^{c\hs{0.5mm}'})\sim (-, +)~,
\la{16par}
\eeq
and opposite parities for the corresponding mirrors, taking into account
(\ref{VSipar}), one can verify that $Z_2\tm Z_2'$ is a
symmetry of the whole 5D Lagrangian. According to (\ref{16par}), at $y=0$
we have the zero modes of $F$, $F^c$ states, which effectively
constitute an anomaly free $16$-plet of $SO(10)$.
As far as the higgs sector is concerned, the 
breaking of $G_{422}$ to MSSM proceeds
through the VEVs of $H^c$, $\ov{H}^c$ states, where the transformation
$H^c$ 
under $G_{422}$ is indicated in (\ref{16dec}).
The zero modes of these states are obtained by introducing 
the supermultiplets
${\bf 16}^{H^c}_{N=2}=(16^{H^c}, \ov{16}^{H^c})$,
${\bf 16}^{\ov{H}^c}_{N=2}=(16^{\ov{H}^c}, \ov{16}^{\ov{H}^c})$, where
\beq
16^{H^c}=H+H^c~,~~\ov{16}^{H^c}=\ov{H}+\ov{H}^{c\hs{0.5mm}'}~,~~
16^{\ov{H}^c}=H'+H^{c\hs{0.5mm}'}~,~~
\ov{16}^{\ov{H}^c}=\ov{H}'+\ov{H}^c~.
\la{16Hdec}
\eeq  
With the parity assignments
\beq
(H^c, \ov{H}^c)\sim (+, +)~,~~(H, \ov{H}')\sim (-, +)~,~~
(H' ,\ov{H})\sim (+, -)~,~~(H^{c\hs{0.5mm}'} ,
\ov{H}^{c\hs{0.5mm}'})\sim (-, -)~,
\la{16Hpar}
\eeq
the $H^c$, $\ov{H}^c$ will contain zero modes.

As far as the bi-doublet $h$ (containing the MSSM higgs doublets)
of $G_{422}$ is concerned, 
it comes from the $10$-plet of $SO(10)$. In 5D we
introduce ${\bf 10}^h_{N=2}=(10^h ,\ov{10}^h)$, 
${\bf 10}^D_{N=2}=(10^D , \ov{10}^D)$, where in terms of $G_{422}$,
\beq
10^h=h(1, 2, 2)+D'(6, 1, 1)~,~~~~
10^D=h'(1, 2, 2)+D(6, 1, 1)~,
\la{10hDdec}
\eeq
while $\ov{10}^h$, $\ov{10}^D$ have conjugate decompositions. With 
$Z_2\tm Z_2'$ parity assignments
\beq
(h, D)\sim (+, +)~,~~~~(h', D')\sim (-, +)~,
\la{10hDpar}
\eeq
and opposite parities for the corresponding mirrors, at the $y=0$ brane 
(fixed point) we find non-vanishing zero modes from the $h$, $D$ states.

It turns out that the $G_{422}$ model with minimal field content encounters
difficulties in realizing a realistic fermion sector. To cure this
problem, following  \cite{our422}, we introduce in the bulk
three families of 
${\bf 10}^f_{N=2}=(10^f, \ov{10}^f)$, 
${\bf 10}^g_{N=2}=(10^g, \ov{10}^g)$ states where, under the $G_{422}$,
\beq
10^f=f(1, 2, 2)+g'(6, 1, 1)~,~~~
10^g=f'(1, 2, 2)+g(6, 1, 1)~.
\la{10fgdec}
\eeq
With parities
\beq
(f, g)\sim (+, +)~,~~~~(f', g')\sim (-, +)~ 
\la{10fgpar}
\eeq
and opposite parities for corresponding mirrors, three families of $f$ and
$g$ have zero mode states.

In summary, we have shown how zero modes of desired $G_{422}$ 
representations
can be obtained from the bulk supermultiplets. Indeed, one can introduce
the needed zero modes directly at the $y=0$ fixed point. For building the
fermion sector and studying its phenomenology the 4D superpotential terms
which we consider below are important. The heavy bulk states do not
affect our conclusions which are therefore robust. 
Finally, the $G_{422}$ representations which contain zero modes and
with which we will deal below are the matter states:
\beq
F_{\al }(4, 2, 1)~,~~F^c_{\al }(\bar 4, 1, 2)~,~~
f_{\al }(1, 2, 2)~,~~g_{\al }(6, 1, 1)~,
\la{422fer}
\eeq
($\al $ is the family index), and the 'scalar' supermultiplets:
\beq
H^c(\bar 4, 1, 2)~,~~\ov{H}^c(4, 1, \bar 2)~,~~h(1, 2, 2)~,~~
D(6, 1, 1)~.
\la{422higgs}
\eeq

\section{Model At Fixed Point $y=0$}

Five dimensional SUSY does not allow Yukawa and higgs
superpotential couplings in the bulk. Because of this, we will construct
the 4D theory at the fixed point $y=0$.
We employ the 4D superfield notation after appropriate 
rescaling from 5D fields.

The breaking of $G_{422}$ to $G_{321}$ can occur through the $H^c$, 
$\ov{H}^c$ states of (\ref{422higgs}). These states contain the 
MSSM singlets
$\nu^c_H$, $\ov{\nu }^c_H$, whose non-zero VEVs break the
$G_{422}$ to the $G_{321}$ \cite{king}. In fact, these VEVs break 
$SU(4)_c\tm SU(2)_R$ to $SU(3)_c\tm U(1)_Y$, where $SU(3)_c$ is a subgroup of
$SU(4)_c$, while $U(1)_Y$ is a superposition of two $U(1)$ factors coming
from $SU(4)_c$ and $SU(2)_R$
\beq
Y=-\sqrt{\fr{2}{5}}Y_{SU(4)_c}+\sqrt{\fr{3}{5}}Y_{SU(2)_R}~,
\la{supU1s}
\eeq
where $Y_{SU(4)_c}$ and $Y_{SU(2)_R}$ are generators of $SU(4)_c$ and
$SU(2)_R$ respectively
\beq
Y_{SU(4)_c}=\fr{1}{\sqrt{24}}{\rm Diag}(1, 1, 1, -3)~,~~~
Y_{SU(2)_R}=\fr{1}{2}{\rm Diag}(1, -1)~.
\la{Y42}
\eeq
In (\ref{supU1s}), the $Y$ hypercharge is given in the 'standard'
$SU(5)$ normalization
\beq
Y=\fr{1}{\sqrt{60}}{\rm Diag}(2, 2, 2, -3, -3)~.
\la{Yhyp}
\eeq
Another superposition, orthogonal to (\ref{supU1s}), is broken by
$\lan \nu^c_H\ran =\lan \ov{\nu }^c_H\ran \equiv v$.
In this breaking, nine degrees of freedom from the scalars are
genuine Goldstone fields that are
absorbed by the appropriate gauge fields. These nine degrees
come from  $\tilde{u}^c$, $\tilde{e}^c$ (tilde indicates scalar
components) and $\tilde{\ov{u}}^c$, $\tilde{\ov{e}}^c$
states (from $H^c$ and $\ov{H}^c$ respectively), plus one superposition
of singlets $\tilde{\nu }^c_H$, $\tilde{\ov{\nu }}^c_H$. Since these states
are complex, there remain nine physical scalars which acquire masses
through the D-terms of $SU(4)_c\tm SU(2)_R$. Their fermionic superpartners
acquire
masses through mixing with appropriate gauginos after 
$G_{422}\to G_{321}$
breaking. In this way, the supersymmetric higgs mechanism is realized
within the $G_{422}$ model.
As far as the physical colored triplet $d^c_H$,
$\ov{d}^c_H$ states are concerned, through 
the couplings $H^cH^cD$, $\ov{H}^c\ov{H}^cD$,
taking into account that $D=(d^c_D ,\ov{d}^c_D)$, and after 
substituting the 
$\lan \nu^c_H\ran $, $\lan \ov{\nu }^c_H\ran $ VEVs, they decouple with 
$\ov{d}^c_D$, $d^c_D$ forming massive states 
$v d^c_H\ov{d}^c_D$,
$v \ov{d}^c_H d^c_D$ \cite{king}.

\subsection{Charged Fermion Masses And Mixings}

The three families of $F_{\al }$, $F^c_{\al }$ states in
(\ref{422fer}) constitute the minimal matter content of $G_{422}$. The
Yukawa couplings for the quark and lepton masses are
$Y^{\al \bt }F_{\al }F^c_{\bt }h$. Recalling that
\beq
F=(q, l)~,~~~F^c=(u^c, d^c, e^c, \nu^c)~, ~~~h=(h_u, h_d)~, 
\la{FFcdec}
\eeq
these couplings lead to phenomenologically
unacceptable asymptotic relations
$\hat{Y}_u^{(0)}=\hat{Y}_d^{(0)}=\hat{Y}_e^{(0)}$, 
$\hat{V}_{CKM}={\rm 1}$. To avoid this, we extend the matter sector of 
$G_{422}$ by invoking the
states $(f+g)_{\al }$ of (\ref{422fer}).
The content of the latter reads
\beq
f=(l_f, \ov{l}_f)~,~~~~g=(d^c_g, \ov{d}^c_g)~.
\la{fgdec}
\eeq
This extension is important for building the desired
charged fermion sector \cite{our422}, and is also crucial for 
the realization of
neutrino democracy. We further introduce a ${\cal U}(1)$ flavor
symmetry for help in realizing the desired
hierarchies. {}To break ${\cal U}(1)$  we introduce an $SO(10)$ singlet
superfield $X$ with charge $Q_X=-1$. We assume that the scalar component
of $X$ develops a VEV
\beq
\fr{\lan X\ran }{M_f}\equiv \ep \simeq 0.2~,
\la{Xvev}
\eeq
where $M_f$ is the cut-off scale of the theory.

In Table \ref{t:charges} we present the ${\cal U}(1)$ charges 
for the various states, where $m$, $n$, $p$, $q$, $k$ are some 
non negative integers, to be specified below.
The charges of $F$, $F^c$ states are fixed from 
the values of CKM mixing angles and hierarchies between the up-type quark 
masses. 
\begin{table} \caption{${\cal U}(1)$ charges of matter and scalar
superfields.}
 
\label{t:charges} $$\begin{array}{|c|c|c|c|c|c|}
 
\hline 
\vspace{-0.4cm}  
&&&&& \\  
\vspace{-0.4cm}   
Q[F_1] &Q[F_2]  &Q[F_3]  &Q[F^c_1] &Q[F^c_2]  &Q[F^c_3] \\

&&&&& \\  
 
\hline
\vspace{-0.4cm}  
&&&&& \\
\vspace{-0.4cm}
\fr{2m-p+q}{4}+3 &\fr{2m-p+q}{4}+2  &\fr{2m-p+q}{4}  & 
\fr{2n-p+q}{4}+5 &\fr{2n-p+q}{4}+2 &\fr{2n-p+q}{4}  \\
 
&&&&& \\ 

\hline \hline

\vspace{-0.4cm}  
&&&&& \\  
\vspace{-0.4cm}    
Q[g_1] &Q[g_2]=Q[g_3]  &Q[f_{\alpha }]=-Q[h]-k  &Q[H^c] &
Q[\overline{H}^c] &Q[D] \\
 
&&&&& \\  
 
\hline
\vspace{-0.4cm} 
&&&&& \\ 
\vspace{-0.4cm}  
\fr{2n-p+q}{2}-k+1 &\fr{2n-p+q}{2}-k   &\fr{m+n-p+q}{2}-k  &
\fr{3p-3q-2n}{4}+k  &\fr{p-q-2n}{4}+k  &\fr{3q+2n-p}{2}-2k \\

&&&&& \\ 
 
\hline 
 
\end{array}$$
 
\end{table} 
With these assignments,
the couplings responsible for the generation of up-type quark masses 
are given by
\begin{equation}
\begin{array}{ccc}
 & {\begin{array}{ccc}
\hspace{-7mm} ~~F^c_1~ & \,\,F^c_2 ~ & \,\,F^c_3 
  
\end{array}}\\ \vspace{2mm}
\begin{array}{c}
F_1 \\ F_2 \\F_3
 \end{array}\!\!\!\!\! &{\left(\begin{array}{ccc}
\,\,\epsilon^8~~ &\,\,\epsilon^{5}~~ &
\,\,\ep^3
\\  
\,\,\epsilon^7~~   &\,\,\epsilon^4~~  &
\,\,\ep^2
 \\
\,\,\epsilon^5~~ &\,\,\epsilon^2~~ &\,\,1
\end{array}\right)h }~,
\end{array}  \!\!  ~~~~~
\label{upY}
\eeq
which, upon diagonalization, yield the desirable 
hierarchies in (\ref{ulam}).

Turning to the down quark and charged lepton sector we will need the
$g_{\al }$ and $f_{\al }$ states. From (\ref{fgdec}), $g$ and
$f$ contain fields with the quantum numbers of $d^c$ and $l$. 
With the ${\cal U}(1)$ charge prescriptions of Table \ref{t:charges}, 
the couplings are 
\begin{equation}
\begin{array}{ccc}
 & {\begin{array}{ccc}
\hspace{-11mm} \,\,g_1~~ &\,\,g_2 ~ & \,\,g_3 
  
\end{array}}\\ \vspace{2mm}
\begin{array}{c}
F^c_1 \\ F^c_2 \\F^c_3
 \end{array}\!\!\!\!\! &{\left(\begin{array}{ccc}
\,\,\epsilon^6~~ &\,\,\epsilon^{5}~~ &
\,\,\ep^5
\\  
\,\,\epsilon^3~~   &\,\,\epsilon^2~~  &
\,\,\ep^2
 \\
\,\,\epsilon ~~ &\,\, 1~~ &\,\,1
\end{array}\right)\ep^nH^c }~.
\end{array}  \!\!  ~~~~~
\label{massD}
\eeq
After substitutution of $\lan H^c\ran $, this ensures the decoupling of
$d^c_{F^c}$ with $\ov{d^c}_g$, so that the light $d^c_{\al }$ states 
reside in $g_{\al }$. With the couplings
\begin{equation}
\begin{array}{ccc}
 & {\begin{array}{ccc}
\hspace{-11mm} \,\,g_1~ & \,\,g_2 ~ & \,\,g_3 
  
\end{array}}\\ \vspace{2mm}
\begin{array}{c}
F_1 \\ F_2 \\F_3
 \end{array}\!\!\!\!\! &{\left(\begin{array}{ccc}
\,\,\epsilon^4~~ &\,\,\epsilon^{3}~~ &
\,\,\ep^3
\\  
\,\,\epsilon^3~~   &\,\,\epsilon^2~~  &
\,\,\ep^2
 \\
\,\,\epsilon ~~ &\,\, 1~~ &\,\,1
\end{array}\right) \fr{\ov{H}^c}{M_f}h }~,
\end{array}  \!\!  ~~~~~
\label{downY}
\eeq
the hierarchies in (\ref{dlam}) can be realized. Since the 
left-handed 
quark doublets $q_{\al }$ reside in $F_{\al }$, from
(\ref{upY}), (\ref{downY}), one can also expect that the
desired values in (\ref{ckm}) for CKM matrix elements can be realized.

As far as charged leptons are concerned, with the  
charge assignments in Table \ref{t:charges}, the
$F_{\al }f_{\bt }H^c$ type couplings have the structure
\begin{equation}
\begin{array}{ccc}
 & {\begin{array}{ccc}
\hspace{-11mm} \,\,f_1~~ & \,\,f_2 ~ & \,\,f_3 
  
\end{array}}\\ \vspace{2mm}
\begin{array}{c}
F_1 \\ F_2 \\F_3
 \end{array}\!\!\!\!\! &{\left(\begin{array}{ccc}
\,\,\epsilon^3~~ &\,\,\epsilon^{3}~~ &
\,\,\ep^3
\\  
\,\,\epsilon^2~~   &\,\,\epsilon^2~~  &
\,\,\ep^2
 \\
\,\,\,1~~ &\,\,\,1~~ &\,\,\,1
\end{array}\right)\ep^m H^c }~.
\end{array}  \!\!  ~~~~~
\label{massL}
\eeq
Substituting in (\ref{massL}) the VEV $\lan H^c\ran $,
one
can easily verify that the $l_F$ states  decouple together 
with $\ov{l}_f$.
Therefore, the light left-handed lepton doublets $l_{\al }$ arise
from $f_{\al }$. Since the 
$f_{\al }$  have identical transformation properties under 
${\cal U}(1)$, we see that democracy between the $l$ states is 
realized! The couplings responsible for charged lepton masses are
\begin{equation}
\begin{array}{ccc}
 & {\begin{array}{ccc}
\hspace{-7mm} f_1~~ & \,\,f_2 ~ & \,\,f_3 
  
\end{array}}\\ \vspace{2mm}
\begin{array}{c}
F^c_1 \\ F^c_2 \\F^c_3
 \end{array}\!\!\!\!\! &{\left(\begin{array}{ccc}
\,\,\epsilon^5~~ &\,\,\epsilon^{5}~~ &
\,\,\ep^5
\\  
\,\,\epsilon^2~~   &\,\,\epsilon^2~~  &
\,\,\ep^2
 \\
\,\,1~~ &\,\,1~~ &\,\,1
\end{array}\right)\fr{\ov{H}^c}{M_f}h }~.
\end{array}  \!\!  ~~~~~
\label{eY}
\eeq
Diagonalization of (\ref{eY}) leads to the required hierarchies 
in (\ref{elam}). 

Note that the up quark mass 
matrix (\ref{upY}) selects $Q[F_1^c]-Q[F_3^c]=5$, $Q[F_2^c]-Q[F_3^c]=2$, 
and since hierarchies between the charged lepton masses are 
$\fr{m_e}{m_{\tau }}\sim \ep^5$, $\fr{m_{\mu }}{m_{\tau }}\sim \ep^2$, 
this dictates the democratic selection
$Q[f_1]=Q[f_2]=Q[f_3]$. Thus, we have a natural and selfconsistent 
realization of neutrino democracy.

Summarizing this subsection, with the help of additional $(f+g)_{\al }$ 
states
and ${\cal U}(1)$ flavor symmetry, we have obtained the desired charged
fermion masses and mixings within 5D SUSY $SO(10)$ model with $G_{422}$ 
symmetry realized at the fixed point $y=0$. Below we discuss 
details of the neutrino sector.

\subsection{Democracy For Bi-large Neutrino Mixings}

The right
handed $\nu^c_{\al }$ states come from $F^c_{\al }$ with the Dirac
couplings given by $\nu^c\hat{Y}^{\nu }_Dlh_u$. The latter arise from
(\ref{eY}), 
which is also responsible for the charged lepton masses.
Thus, due to $SU(2)_R$ symmetry which relates $\nu^c $ with $e^c$ and
$h_u$ with $h_d$, we have
\beq
\hat{m}^{\nu }_D=\hat{m}_e\tan \bt ~.
\la{DEuni}
\eeq
This relation allows us to estimate the Majorana mass of the $\nu^c_3$
state which, in turn, is responsible for the generation of the atmospheric 
neutrino scale. The $\nu^c_3$ acquires mass through the coupling
\beq
\ep^{2k}\fr{\lam }{M_f}(F^c_3\ov{H}^c)^2~,
\la{maj3}
\eeq
which gives
\beq
M_R^{(3)}=\fr{\lam \ep^{2k}}{M_f}\lan \ov{H}^c\ran^2~.
\la{maj3mass}
\eeq
Taking into account (\ref{DEuni}), we have
\beq
m_{\nu_3}=\fr{[(\hat{m}^{\nu }_D)_{33}]^2}{M_R^{(3)}}=
\fr{m_{\tau }^2\tan^2 \bt }{M_R^{(3)}}~.
\eeq
With hierarchical masses for neutrinos (which indeed must be the case
for a democratic scenario), we have 
$m_{\nu_3}=\sqrt{\De m_{\rm atm}^2}$. Thus,
\beq
M_R^{(3)}=\fr{m_{\tau }^2\tan^2 \bt }{\sqrt{\De m_{\rm atm}^2}}=
\left\{ \begin{array}{lll}
6\cdot 10^{10}~{\rm GeV}; &~ {\rm for}~\tan \bt \sim 1 \\
6\cdot 10^{12}~{\rm GeV}; &~ {\rm for}~\tan \bt \sim 10 \\
2\cdot 10^{14}~{\rm GeV}; &~ {\rm for}~\tan \bt \sim 60
\end{array} 
\right.~.
\la{MR3pred}
\eeq
For low and intermediate values of $\tan \bt $ the predicted
value of $M_R^{(3)}$ is relatively low
which opens up an interesting
possibility for implementing either inflationary
\cite{infry} or thermal \cite{sacha}  leptogenesis.

Due to $l$-democracy, one could naively expect that 
$m_{\nu_1}\sim m_{\nu_2}\sim m_{\nu_3}$. 
This scale for $\nu_2$ would be inappropriate for resolving the solar
neutrino anomaly. To avoid this, some care should be exercised. One
way is to introduce $SO(10)$ singlet states $N_1$ and $N_2$ with
${\cal U}(1)$ charges $-5-k$ and $-2-k$ respectively. Then, 
with the couplings $(F^c_1N_1+F^c_2N_2)\ov{H}^c$,
after substituting the $\ov{H}^c$ VEV, $\nu^c_1$ and $\nu^c_2$ decouple
together
with $N_1$ and $N_2$. At this stage $m_{\nu_1}=m_{\nu_2}=0$, since 
with the
introduction of $N_{1, 2}$ states, the lepton numbers $L_e$, $L_{\mu }$
are conserved. To generate a mass for $\nu_2$, we introduce an additional
$SO(10)$ singlet ${\cal N}$ with charge $Q({\cal N})=k$. With couplings
\beq
(\lam_1f_1+\lam_2f_2+\lam_3f_3){\cal N}h+M_f\ep^{2k}{\cal N}^2~,
\la{calN}
\eeq
integration of the ${\cal N}$ state can generate the scale relevant 
for the solar neutrino anomaly\footnote{
These $SO(10)$ singlet states can be introduced either directly on the
brane or in the bulk. In the bulk case, the accompanying mirrors carry
opposite orbifold parities and ${\cal U}(1)$ charges. 
The presence of mirrors does not affect results obtained through 
the brane superpotential couplings.}.
The additional singlet states are necessary for generating appropriate
mass scale for solar neutrino and self consistent bi-large neutrino
oscillations.

Let us consider the scheme in some detail. From 
(\ref{DEuni}), the $\nu^c-l$ and $e^c-l$ couplings are diagonalized 
simultaneously. Although the introduction of $N_{1, 2}$ makes 
$\nu^c_{1, 2}$  irrelevant, it is still convenient to work with basis
in which the matrices in (\ref{DEuni}) are diagonal. 
In this case, the lepton mixing 
matrix will coincide with the unitary matrix which diagonalizes neutrino 
mass matrix, with non-trivial neutrino mixings coming from the right 
handed neutrino sector, i.e. from the $\nu^c_3-{\cal N}$ mixing term. 
The ${\cal U}(1)$ symmetry allows the coupling $F^c_3{\cal 
N}\ov{H}^c\ep^{2k}$, 
which gives a mixing term $M_{3{\cal N}}\nu^c_3{\cal N}$, with
$M_{3{\cal N}}\simeq \lan \ov{H}^c\ran \ep^{2k}$. Taking into account all 
this and also (\ref{maj3mass}), (\ref{calN}), the relevant matrix 
is given by 
\begin{equation}
\begin{array}{ccccc}
 & {\begin{array}{ccccc}
\hspace{-7mm} ~~\nu_1~ & ~~\,\nu_2 ~ & \,~~\nu_3~ &~~~ \nu^c_3 &
~~~~{\cal N} 
  
\end{array}}\\ \vspace{2mm}
\begin{array}{c}
\nu_1 \\ \nu_2 \\ \nu_3 \\ \nu^c_3 \\ {\cal N}
 \end{array}\!\!\!\!\! &{\left(\begin{array}{ccccc}
\,\,0~~ &\,\,0~~ &\,\,0~~ & 0 &\lam_1h_u
\\  
\,\,0~~   &\,\,0~~  &\,\,0~~ & 0 &\lam_2h_u
 \\
\,\,0~~ &\,\,0~~ &\,\,0~~ &\,\,\lam_{\tau }h_u & \lam_3h_u
\\
\,\,0~~ &\,\, 0~~ & \lam_{\tau }h_u  &\,\, M_R^{(3)} & M_{\cal N}\ep'
\\
\lam_1h_u & \lam_2h_u & \lam_3h_u & \,\,M_{\cal N}\ep' & M_{\cal N}
\end{array}\right)}~,
\end{array}  \!\!  ~~~~~
\label{bignu}
\eeq
where we have defined $M_{\cal N}=M_f\ep^{2k}$, 
$\ep'=\fr{M_{3{\cal N}}}{M_{\cal N}}\sim \fr{\lan \ov{H}^c\ran }{M_f}$.

{}From (\ref{bignu}) one sees that $\nu^c_3-{\cal N}$ mixing plays 
an essential role. Namely, to have large $\nu_{\mu }\to \nu_{\tau }$  
oscillations, one should have $\lam_{2, 3}\ep'\sim \lam_{\tau }$. Having 
in mind that $\ep'\ll 1$, we also have $\lam_{\tau }\ll 1$, which
means that $\tan \bt $ is not large, but has either a low or an intermediate 
value. Integrating out ${\cal N}$ and $\nu^c_3$ in 
(\ref{bignu}), we obtain the light neutrino mass matrix
\begin{equation}
\begin{array}{ccc}
 & {\begin{array}{ccc}
~ & &\,\,~~~
\end{array}}\\ \vspace{2mm}
\begin{array}{c}
 \\  \\ 

\end{array}\!\!\!\!\! &\hat{m}_{\nu }={\left(\begin{array}{ccc}
\lam_1^2& \lam_1\lam_2&
\lam_1\ov{\lam }_3
\\
\lam_1\lam_2 &\lam_2^2&
\lam_2\ov{\lam }_3
\\
\lam_1\ov{\lam }_3&
\lam_2\ov{\lam }_3&
\ov{\lam }_3^2
\end{array}\right)m }~+
\end{array}  
\hs{-1cm}
\begin{array}{ccc}
& {\begin{array}{ccc}
 &  & 
\end{array}}\\ \vspace{2mm}
\begin{array}{c}
  \\  \\

\end{array} &{\left(\begin{array}{ccc}
\lam_1^2& \lam_1\lam_2 & \lam_1\lam_3
\\
\lam_1\lam_2  & \lam_2^2 & \lam_2\lam_3 
\\
\lam_1\lam_3  & \lam_2\lam_3&  \lam_3^2
\end{array}\right)m'~,
}
\end{array}
\label{smallnu}
\end{equation}                
where 
\beq
m=\fr{\ep'^2h_u^2}{M_R^{(3)}}~,~~~
m'=\fr{h_u^2}{M_{\cal N}}~,~~~
\ov{\lam}_3=\lam_3-\lam_{\tau }/\ep'~.
\la{scales}
\eeq
The scales $m$, $m'$ correspond to atmospheric and solar neutrino 
anomalies, so $m\gg m'$.

For analyzing the neutrino mass matrix, it is convenient to rewrite 
(\ref{smallnu}),
\begin{equation}
\begin{array}{ccc}
 & {\begin{array}{ccc}
\hspace{-7mm}  & & 
  
\end{array}}\\ \vspace{2mm}
\begin{array}{c}
 \\  \\
 \end{array}&\hat{m}={\left(\begin{array}{ccc}
\,\,\tl{\lam }_1^2~~ &\,\,\tl{\lam }_1\tl{\lam }_2~~ &
\,\, \tl{\lam }_1\tl{\lam }_3
\\  
\,\,\tl{\lam }_1\tl{\lam }_2~~   &\,\,\tl{\lam }_2^2~~  &
\,\,\tl{\lam }_2\tl{\lam }_3
 \\
\,\,\tl{\lam }_1\tl{\lam }_3~~ &
\,\,\tl{\lam }_2\tl{\lam }_3~~ &\,\,\tl{\lam }_3^2(1+\de )
\end{array}\right)m }~,
\end{array}  \!\!  ~~~~~
\label{rednu}
\eeq
where
\beq
\tl{\lam }_{1, 2}=\sqrt{1+m'/m}\lam_{1, 2}~,~~~
\tl{\lam }_3=
\fr{1+\lam_3m'/(\ov{\lam }_3m)}{\sqrt{1+m'/m}}
\ov{\lam }_3~,~~~ 
\de =(1-\fr{\lam_3}{\ov{\lam }_3}+\fr{\lam_3^2}{\ov{\lam }_3^2})
\fr{m'}{m}~.
\la{lamtil}
\eeq 
The value of $\de $ is small and is responsible for the non-zero mass 
of $\nu_2$. From (\ref{rednu}), 
assuming $\lam_{\al }\sim \lam_{\tau }/\ep'$ one obtains 
\beq
m_{\nu_1}=0~,~~~
m_{\nu_2}\simeq \fr{(\tl{\lam }_1^2+\tl{\lam }_2^2)\tl{\lam }_3^2}
{\tl{\lam }_1^2+\tl{\lam }_2^2+\tl{\lam }_3^2} \de m 
\sim \tl{\lam }_{\al }^2m'~,
m_{\nu_3}\simeq (\tl{\lam }_1^2+\tl{\lam }_2^2+\tl{\lam }_3^2)m~.
\la{massnu}
\eeq
As expected, the neutrino masses are hierarchical, with 
$\De m^2_{atm }\sim m_{\nu_3}^2$, $\De m^2_{sol }\sim m_{\nu_2}^2$.
From (\ref{scales}), (\ref{massnu}), we see that 
the relevant scale for the solar neutrino anomaly
$m_{\nu_2}\sim 10^{-2}$~eV
is generated for 
$M_{\cal N}\simeq (3\cdot 10^{11} \tan^2 \bt/\ep'^2)$~GeV.
With $\ep'\simeq 0.2 $ and $\tan \bt =1-60$,
$M_{\cal N}\simeq (10^{13}-3\cdot 10^{16})$~GeV.
For the atmospheric neutrinos, 
$m_{\nu_3 }\simeq 5\cdot 10^{-2}$~eV is generated with the  
$M_R^{(3)}$ scales indicated in (\ref{MR3pred}).   

As far as the mixing angles are concerned, from the structure of 
(\ref{rednu}), (\ref{smallnu}), with 
$\lam_1\sim \lam_2\sim \lam_3\sim \lam_{\tau }/\ep' $, taking into 
account (\ref{scales}), (\ref{lamtil}), one naturally expects large neutrino 
mixings \beq
\sin^2 2\te_{\mu \tau }\sim 1~,~~~\sin^2 2\te_{e \mu, \tau }\sim 1~,
\la{lepmix}
\eeq
as intended.

As was emphasized in \cite{ourdem}, \cite{5Dourdem}, within the democratic 
approach, one also expects a large value for the $\te_{13}$ 
mixing angle. On the other hand, the CHOOZ data \cite{chooz} requires
$\te_{13}\stackrel{<}{_\sim }0.2(\simeq \ep)$, so that some  
cancellation among the parameters is
needed.  If future measurements turn out to favor a much smaller
($\ll \ep $) 
$\te_{13}$, some modification of the demoratic approach would 
be required.

\subsection{Particle Spectra And Gauge Coupling Unification}

Below the symmetry breaking scale of $G_{422}$ ($v\simeq M_G$), the massless 
gauge fields are just those from the MSSM. However, there are some
additional vector like supermultiplets with masses below the GUT scale 
$M_G$. From (\ref{massD}) and
(\ref{massL}), it is easy to see that the masses of decoupled color triplets
and $SU(2)_L$ doublet states respectively are
$$
m_{t_1}\simeq v\ep^{n+6}~,~~~~m_{t_2}\simeq v\ep^{n+2}~,~~~~
m_{t_3}\simeq v\ep^{n}~,
$$
\beq
m_{d_1}\simeq v\ep^{m+3}~,~~~~m_{d_2}\simeq v\ep^{m+2}~,~~~~
m_{d_3}\simeq v\ep^{m}~.
\la{ferDTmas}
\eeq
With the ${\cal U}(1)$ charges in Table \ref{t:charges}, the superpotential
couplings
\beq
W(H, D)=\ep^p H^cH^cD+\ep^q\ov{H}^c\ov{H}^cD~,
\la{supHD}
\eeq
yield masses 
\beq
m_{T_1}\simeq v\ep^p~,~~~~~~~~m_{T_2}\simeq v\ep^q~,
\la{scDTmas}
\eeq
for the colored triplet states from 'scalar'
supermultiplets.

In the  simplest case, one can assume that the compactification 
scale $\mu_0$ and the
$G_{422}$ breaking scale  are close to  
$M_G$($\simeq \mu_0\simeq v$). On the other hand,
from (\ref{ferDTmas}), (\ref{scDTmas}) we see that 
with $m, n, p, q\neq 0$, 
below the scale $v$ there are additional states which contribute, 
together with the
MSSM states, to the renormalization of the gauge couplings.
For the strong coupling constant in particular,
\beq
\al_3(M_Z)^{-1}=[\al_3(M_Z)^{-1}]^{\rm min}_{\small SU(5)}+\fr{9}{14\pi }
\ln \fr{M_G^2m_{d_1}m_{d_2}m_{d_3}}{m_{t_1}m_{t_2}m_{t_3}m_{T_1}m_{T_2}}~,
\la{als}
\eeq
where the first term on the right hand side corresponds to the  
value
calculated in minimal SUSY $SU(5)$ [MSSU(5)]. In order to preserve 
the MSSU(5) prediction, one should take
$m_{t_1}m_{t_2}m_{t_3}m_{T_1}m_{T_2}\simeq M_G^2m_{d_1}m_{d_2}m_{d_3}$, which
taking into account (\ref{ferDTmas}), (\ref{scDTmas}) gives the relation
\beq
3n+p+q+3=3m~.
\la{constr}
\eeq
If in the right hand side of (\ref{constr}), $3m$ is replaced by $3m+c$ 
(c=1, 2), one 
would have the possibility to reduce the somewhat large value of 
$\al_3(M_z)$($\simeq 
0.126$) predicted by MSSU5 \cite{al3MSSU5}. Thus, a proper selection 
of the integers $m, n, p, q$ can yield successful unification with 
additional matter states playing a crucial role.

\section{Proton Stability And Automatic Matter Parity}

In this section we  discuss the issue of nucleon decay within our
scheme. Let us start with dimension five left and right handed
operators
\beq
{\cal O}_L=qqql~,~~~~~{\cal O}_R=u^cu^cd^ce^c~,
\la{d5ops}
\eeq
whose presence depends on the details of the model. 
The ${\cal O}_L$ can emerge through $qqT$, $ql\bar T$ type couplings,
while ${\cal O}_R$ can appear through $e^ce^cT$, $u^cd^c\bar T$,
once the color triplets $T$, $\bar T$ are integrated out. 
Consider an ${\cal R}$ symmetry under which
$W\to e^{2{\rm i}\al }W$ and
$\phi_i\to e^{{\rm i}R_i}\phi_i$, where $2\al $ and $R_i$ are the phases of
the superpotential $W$ and $\phi_i$ superfield respectively. It is easy to 
check that the transformations
$$
(F, F^c, g, f, N_{1, 2},~ {\cal N})\to e^{{\rm i}\al }
(F, F^c, g, f, N_{1, 2}, ~{\cal N})~,~~~
$$
\beq
(h, H^c, \ov{H}^c, X)\to (h, H^c, \ov{H}^c, X)~,~~~
D\to e^{2{\rm i}\al }D~,~~~W\to e^{2{\rm i}\al }W~,
\la{Rcharges}
\eeq 
are consistent with the ${\cal R}$-symmetry. Note that the 
direct mass terms $\ov{H}^cH^c$ and $D^2$ in $W(H, D)$ of (\ref{supHD})
are forbidden to all orders. Therefore, the color triplet mass matrix 
is given by
\begin{equation}
\hat{M}_T=
\begin{array}{cc}
 & {\begin{array}{cc}

\hspace{-7mm} ~~~\ov{d}^c_H~ & \,\,~~\ov{d}^c_D   
  
\end{array}}\\ \vspace{2mm}
\begin{array}{c}
d^c_H \\ d^c_D 
 \end{array}\!\!\!\!\! &{\left(\begin{array}{cc}
\,\,0~~ &\,\,v\ep^p~~ 
\\  
\,\,v\ep^q~~   &\,\,0~~
\end{array}\right) }~.
\end{array}  \!\!  ~~~~~
\label{scTmatr}
\eeq 
{}From (\ref{scTmatr}), the elements of the inverse mass matrix 
which are relevant for nucleon decay, read
\beq
\l \hat{M}_T^{-1}\r_{11}=\l \hat{M}_T^{-1}\r_{22}=0~,~~~ 
\l \hat{M}_T^{-1}\r_{12}=\fr{1}{v\ep^q}~,~~~ 
\l \hat{M}_T^{-1}\r_{21}=\fr{1}{v\ep^p}~. 
\la{invMt}
\eeq 
Thus, for nucleon stability it is crucial that the triplets from the $D$
state do not couple with the matter fields. It is easy to check, using
(\ref{Rcharges}), that the couplings
\beq
FFD~,~~~FfD\ov{H}^c~,~~~F^cgD\ov{H}^c~,~~~F^cF^cD~,
\la{Dfer}
\eeq
are forbidden. Thus,
$d=5$ operators from the color triplet exchange are absent.  

As far as
the Planck scale $d=5$ operators of (\ref{d5ops}) are concerned, one can
also verify that the couplings
\beq
FFFf\ov{H}^c~,~~~~~F^cF^cF^cg\ov{H}^c~,
\la{d5plank}
\eeq
from which these operators could emerge, are also forbidden. 

In principle, in (\ref{scTmatr}) instead of zeros, entries of the 
order of 
SUSY breaking scale $m_S\sim 1$~TeV can be expected. The latter can come 
from the K\"ahler potential after SUSY is broken. By the same reasoning, 
operators (\ref{Dfer}), (\ref{d5plank}), suppressed by $m_S/M_f$, may 
also emerge. However, proton decay is then 
strongly suppressed and essentially unobservable. We
conclude that through a proper ${\cal R}$-charge selection, 
dimension five nucleon decay is eliminated.

Turning to dimension six nucleon decay, with all matter 
supermultiplets introduced in the
bulk, the 5D bulk kinetic terms are irrelevant for nucleon decay
\cite{arch}. This follows since through the exchange of $V_X$, $V_Y$
bosons, the quark-lepton states are converted into heavy states with
masses of order $1/R$. A different source for $d=6$ nucleon decay can be
brane
localized operators, which respect $G_{422}$ and the orbifold
symmetries. This kind of operator has the form \cite{locop} 
\beq
\de (y)\psi_1^+(\pl_5e^{2\hat{V}}-\hat{\Si }e^{2\hat{V}}-
e^{2\hat{V}}\hat{\Si })\psi_2~,
\la{d6loc}
\eeq
where $\psi_1$ and $\psi_2$ denote appropriate
matter superfields with zero modes, and $\hat{V}$, $\hat{\Si }$ 
are fragments
from the coset $SO(10)/G_{422}$.
In order for these operators to be invariant under ${\cal U}(1)$,
either the multiplier $(X^+)^{Q_1}X^{Q_2}$ or $X^{Q_2-Q_1}$
should be present, where $Q_1$, $Q_2$ 
are the ${\cal U}(1)$ charges of $\psi_1$, $\psi_2$. 
If either $Q_1$, $Q_2$  or $Q_2-Q_1$ is not an integer,
the corresponding operator is not allowed.
The operators $F^+\pl_5V_BF^c$,
$f^+\pl_5V_Bg$ carry ${\cal U}(1)$ charge $(n-m)/2$ and will not be
allowed
for $n-m=2k'+1$ ($k'$ is an integer). Thus, with the help of 
${\cal U}(1)$ symmetry, dimension six nucleon decay can also be 
avoided \cite{5Dourdem} within 5D SUSY
$SO(10)$ GUT. 

Alternatively, if the selection $n-m=2k'$ is made, the 
appropriate operators will be allowed with a suppression 
factor $\ep^{k'}$. 
Together with this, there will appear additional powers of $\ep $, coming 
from the ${\cal U}(1)$ charges of the light families.  
The appropriate amplitude has at least an $\ep^2$ suppression. 
Consequently, the amplitudes are proportional 
to $\ep^{2k'+2}$, which even for $k'=0$, gives an 
enhancement factor $\sim 650$ for the lifetime, in comparision to the 
estimates in \cite{locop}. 
This is well within the current 
experimental bounds \cite{SKpdecay}. It is interesting to note that the 
operator 
$\ep^{2k'+2}q_1l_{\al }u^{c+}_1d^{c+}_2 $, relevant for decay with 
emmision of charged leptons, would give
$\tau (p\to Ke)\sim \tau (p\to K\mu )$. This happens due the fact that 
the $l_{\al }$ states all have the same ${\cal U}(1)$ charge.

To conclude this section, we note that the mass term for the
bidoublet $h$ is forbidden by ${\cal R}$-symmetry and therefore at this
level the $\mu $-term is zero. This gives a good starting point for the 
resolution of SUSY $\mu $ problem, although some mechanism \cite{mu1term},
\cite{mu2term} for its
generation should be applied. The same ${\cal R}$-symmetry also forbids all
matter parity violating operators otherwise allowed by the  
symmetry $G_{422}$. That is,
together with the suppression of $d=5$ baryon number violating operators
and $\mu $-term, the ${\cal R}$-symmetry also ensures automatic matter parity
\cite{RvsBPar}, \cite{our422}.

\section{Rare $l_{\al } \to l_{\bt }\ga $  Decays}

In our scenario neutrino masses are generated by the see-saw mechanism 
and within the SUSY framework, it turns out to be also the dominant 
source 
for lepton flavor violating (LFV) rare processes \cite{LFV} such as
$l_{\al } \to l_{\bt }\ga $. Since the masses of the right handed neutrinos 
are below the GUT scale, they contribute to renormalization so that 
universality (assumed to hold at high scales) among the soft slepton  
masses is
violated. For non-universal contributions, the relevant scales are those  
where the 
corresponding right handed neutrinos decouple.  
It is convenient to work with a basis in which the right handed neutrino 
mass matrix is diagonal. The non-universal contributions to the slepton 
masses are then
\beq
(\de m^2)_{\al \bt }\approx -\fr{1}{8\pi^2}(A+3)m_S^2
\sum_{k}(\tl{Y}_{\nu}^T)_{\al k}
\ln \fr{M_G}{M_R^{(k)}}(\tl{Y}_{\nu })_{\bt k}~,
\la{insert1}
\eeq
where we have assumed universality  and proportionality at $M_G$.
Also, in (\ref{insert1})  SUSY breaking occurs through $N=1$ SUGRA.
The $\tl{Y}_{\nu }$ comes 
from the term $N_R'\tl{Y}_{\nu }lh_u$, and $N_R'$ denotes the right handed 
neutrino 
state in a mass eigenstate basis. From (\ref{bignu}), 
\begin{equation}
\begin{array}{cc}
  {\begin{array}{ccc}
\hspace{-7mm}  & & 
  
\end{array}}\\ \vspace{2mm}
\begin{array}{c}
 \\  \\
 \end{array}&\tl{Y}_{\nu }={\left(\begin{array}{ccc}
\,\, -\lam_1\ep'~~ &\,\, -\lam_2\ep'~~ &\lam_{\tau }-\lam_3\ep'
\,\, 
\\  
\,\, \lam_1~~   &\,\, \lam_2~~  & \,\, \lam_3
\end{array}\right) }~.
\end{array}  \!\!  ~~~~~
\label{YnuR}
\eeq
From (\ref{insert1}), (\ref{YnuR}), taking into account that $\ep'\ll 1$, 
one finds
\begin{equation}
\begin{array}{ccc}
 & {\begin{array}{ccc}
\hspace{-7mm}  & & 
  
\end{array}}\\ \vspace{2mm}
\begin{array}{c}
 \\  \\
 \end{array}&(\de m^2)_{\al \bt }\approx -\fr{1}{8\pi^2}(A+3)m_S^2
{\left(\begin{array}{ccc}
\,\, \lam_1^2~~ &\,\, \lam_1\lam_2~~ &
\,\, \lam_1\lam_3
\\  
\,\, \lam_1\lam_2~~   &\,\, \lam_2^2~~  &
\,\, \lam_2\lam_3
 \\
\,\, \lam_1\lam_3~~ &
\,\, \lam_2\lam_3~~ &\,\, \lam_3^2
\end{array}\right)_{\al \bt}\cdot \ln \fr{M_G}{M_{\cal N}}
}~.
\end{array}  \!\!  ~~~~~
\label{insert2}
\eeq
Thus, the ${\cal N}$ state provides the dominant contribution to LFV. 
Due to democracy, the $\lam_{\al }$ couplings all have similar values, 
and therefore one can expect
\beq
{\rm BR}(\mu \to e\ga )\sim {\rm BR}(\tau \to \mu \ga )
\sim {\rm BR}(\tau \to e\ga )~.
\la{demBr}
\eeq
This is a characteristic signature of the democratic scenario. 

More 
precisely, taking into account that $\lam_{\al }\sim \lam_{\tau }/\ep'$, 
eq. (\ref{insert2}) and the expressions given in \cite{BRrare}, 
the branching ratios are 
\beq
{\rm BR}(l_{\al }\to l_{\bt }\ga )=
\fr{\al_{em}^3[(\de m^2)_{\al \bt }]^2}{G_F^2m_S^8}\tan^2 \bt \sim
\fr{\al_{em}^3(A+3)^2}{64\pi^4G_F^2m_S^4}
\l \fr{\lam_{\tau }}{\ep'}\r^4\tan^2 \bt \ln^2 \fr{M_G}{M_{\cal N}}~.
\la{Brs}
\eeq 
In order to satisfy the most stringent bound 
${\rm BR}(\mu \to e\ga)\stackrel{<}{_\sim }10^{-14}$ \cite{BR}, for 
$m_S\sim 1$~TeV, [and if there is no cancellation in (\ref{Brs})],
one should take $\lam_{\tau }/\ep'\sim 5\cdot 10^{-2}$.
With 
$\ep'\sim 0.2$ this gives $\lam_{\tau }\sim 10^{-2}$, 
indicating a preference for low $\tan \bt $ regime, which is also 
suggested by the charged fermion and neutrino sectors.

In conclusion, let us note that in contrast to previously discussed LFV 
scenarios \cite{modelsRare}, our scheme predicts
'democracy' in the sector of LFV rare processes which hopefully can be 
probed in the near future.

\section{Conclusions}

We have presented a realistic model of charged fermion and neutrino
masses and mixings in which a number of key ideas play an essential role.
Namely, $SO(10)$ grand unification, an extra dimension, orbifold and higgs 
breaking and an abelian ${\cal U}(1)$ flavor symmetry. Some additional 
states are 
introduced in order to realize a democratic approach in the neutrino sector,
while yielding the CKM mixings in the quark sector. Note that it is possible
to replace ${\cal U}(1)$ with a discrete symmetry, or with a vectorlike 5D 
gauge symmetry \cite{ourBL}. The issue of proton stability is resolved 
and interesting predictions for rare leptonic decays have been obtained.



\vs{0.2cm} 
\hs{-0.6cm}{\bf Acknowledgments} 

\hs{-0.6cm}We acknowledge the support of NATO Grant
PST.CLG.977666. This work is partially supported  by DOE under
contract DE-FG02-91ER40626.
Z.T. would like to thank the Bartol Research Institute for
warm hospitality during his visit there.



\bibliographystyle{unsrt}

\end{document}